\documentclass[preprint,showpacs,preprintnumbers,amsmath,amssymb,superscriptaddress,prb]{revtex4}

\usepackage{graphicx,amsmath,amsfonts,amssymb}
\usepackage{dcolumn}
\usepackage{bm}
\usepackage{color}

\begin{document}

\title{The Importance of Edge Effects on the Intrinsic Loss Mechanisms of Graphene Nanoresonators}

\author{Sung Youb Kim}
    \affiliation{Department of Mechanical Engineering, University of Colorado, Boulder, CO 80309}
\author{Harold S. Park\footnote{Electronic address: harold.park@colorado.edu}}
    \affiliation{Department of Mechanical Engineering, University of Colorado, Boulder, CO 80309}

\date{\today}

\begin{abstract}

We utilize classical molecular dynamics simulations to investigate the intrinsic loss mechanisms of monolayer graphene nanoresonators undergoing flexural oscillations.  We find that spurious edge modes of vibration, which arise not due to externally applied stresses but intrinsically due to the different vibrational properties of edge atoms, are the dominant intrinsic loss mechanism that reduces the Q-factors.  We additionally find that while hydrogen passivation of the free edges is ineffective in reducing the spurious edge modes, fixing the free edges is critical to removing the spurious edge-induced vibrational states.  Our atomistic simulations also show that the Q-factor degrades inversely proportional to temperature; furthermore, we also demonstrate that the intrinsic losses can be reduced significantly across a range of operating temperatures through the application of tensile mechanical strain.

\end{abstract}


\maketitle

\section{Introduction} 

Graphene has recently been discovered as the simplest two-dimensional crystal structure~\cite{novoselovNATURE2005,novoselovPNAS2005}.  Since then, graphene has been intensely studied for its unusual electrical properties~\cite{geimNM2007}, for its potential in nanocomposites~\cite{stankovichNATURE2006}, and also for its potential as the basic building block of future nanoelectromechanical systems (NEMS)~\cite{bunchSCIENCE2007,frankJVSTB2007,sanchezNL2008,robinsonNL2008a,leeSCIENCE2008}.

As a building block for NEMS, graphene shows particular promise for ultrasensitive detection of masses, forces and pressure due to its own extremely low mass.  However, the key issue limiting the applicability of graphene as a sensing component is its extremely low quality (Q)-factor; the Q-factors of a 20-nm thick multilayer graphene sheet were found to range from 100 to 1800 as the temperature decreased from 300 K to 50 K~\cite{bunchSCIENCE2007}.  Recently, using multilayer graphene oxide films, higher Q-factors with values up to 4000 were found~\cite{robinsonNL2008a}.  A higher Q-factor is critical to NEMS device performance and reliability as it implies less energy dissipation per vibrational cycle, which enables the graphene NEMS to extend its operational lifetime by performing near optimal capacity for a longer period of time.  Furthermore, because the mass or force sensing resolution is inversely proportional to Q~\cite{stoweAPL1997,ekinciRSI2005}, low Q-factors are the key limiting factors to the development of highly sensitive and reliable graphene-based NEMS.  

A recent experimental study by Sanchez \emph{et al.}~\cite{sanchezNL2008} studied the oscillations of suspended multilayer graphene sheets in which two edges of the graphene sheet were fixed, while the two other free edge surfaces of graphene remained free to oscillate.  The Q-factors measured in that work were extremely small, between 2 and 30; the low Q's were attributed to extrinsic loss mechanisms such as air damping effects.  

Interestingly, the study also found that the free edges of the graphene sheet often had the largest vibrational amplitudes during resonance.  To verify this, finite element simulations of the graphene sheet were performed; by introducing non-uniform stresses in the suspended graphene sheet through application of both an in-plane stretch and an in-plane rotation, the authors were able to reproduce the large edge modes of vibration observed experimentally.  However, no correlation between the large edge modes of vibration and the Q-factors were established in that work.

In this letter, we demonstrate using classical molecular dynamics (MD) simulations~\cite{brennerJPCM2002} that spurious edge modes that arise during oscillation from the free edges of the graphene sheet constitute the key \emph{intrinsic} loss mechanism in reducing the Q-factors of oscillating graphene membranes; extrinsic loss mechanisms~\cite{yasumuraJMS2000} such as gas damping and clamping losses are not considered in the present work.  More significantly, we demonstrate that the spurious edge modes arise intrinsically due to the free edges; no externally applied non-uniform stress field is necessary to activate these spurious modes of oscillation.   We also find that passivation of the free edges with hydrogen is ineffective in removing the edge modes, and demonstrate that fixing the edge atoms is the most effective way to reduce the intrinsic edge-induced dissipation.  Finally, we demonstrate the effectiveness of applied tensile mechanical strain in dramatically reducing intrinsic energy dissipation across a range of operating temperatures.

\section{Results}

We first show the results of MD simulations that elucidate the fundamental importance of edge effects in reducing the Q-factors of graphene nanoresonators.  To elucidate the edge effects, we studied a 19.7~\AA$\times$127.8~\AA~graphene monolayer that was comprised of 960 carbon atoms, and constrained both long edges (127.8~\AA) to remain fixed during oscillation.  We utilized three different boundary conditions on the two short (19.7~\AA) edges of the graphene sheet.  First, we applied periodic boundary conditions (PBCs), to eliminate any edge effects resulting from carbon atoms having a reduced number of bonding neighbors.  Second, we let the short edges oscillate freely, with no constraints on their motion.  Third, we passivated the dangling bonds of the short edge carbon atoms with hydrogen.

\begin{figure} \begin{center} 
\includegraphics[scale=0.41]{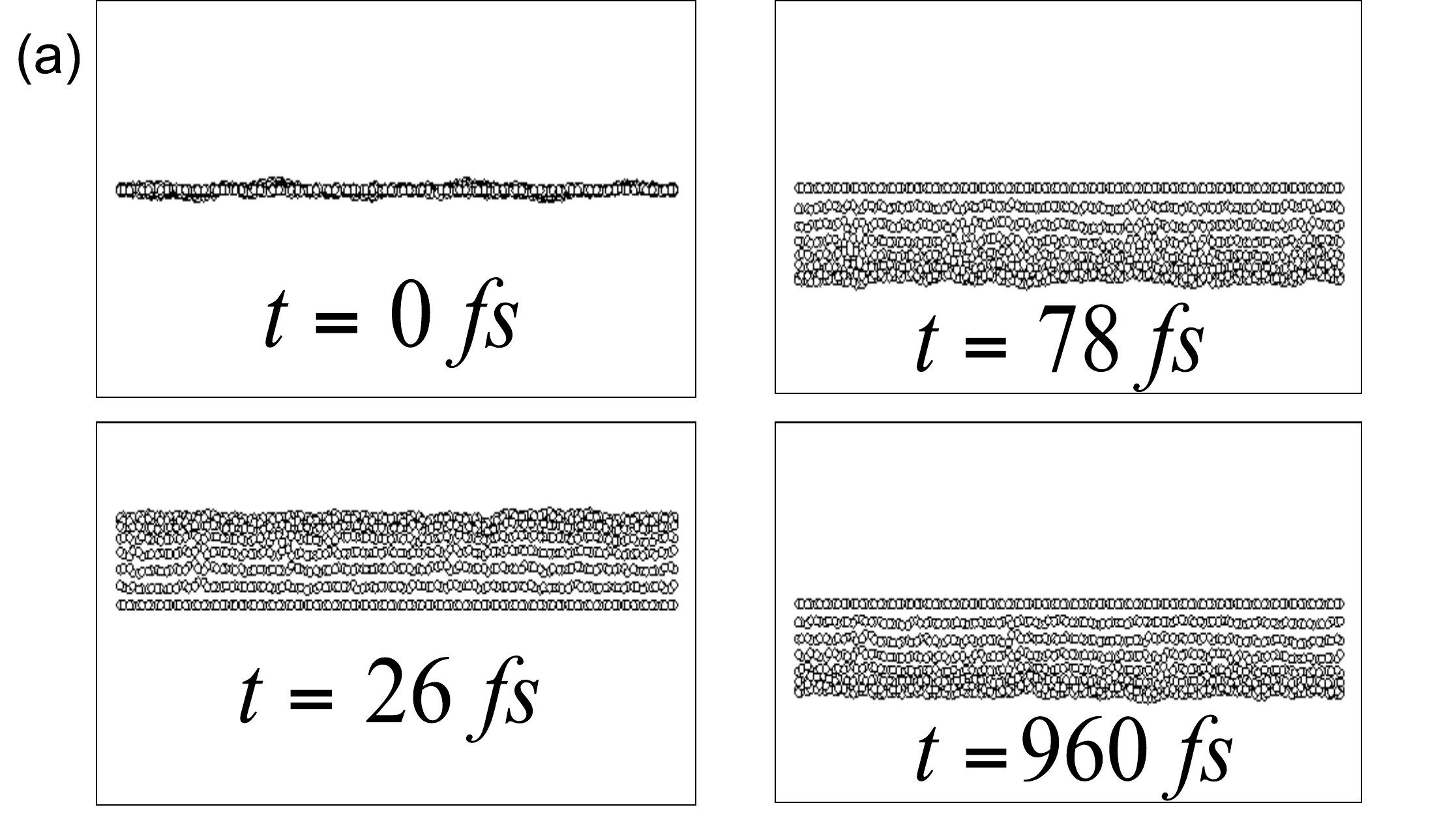}
\includegraphics[scale=0.54]{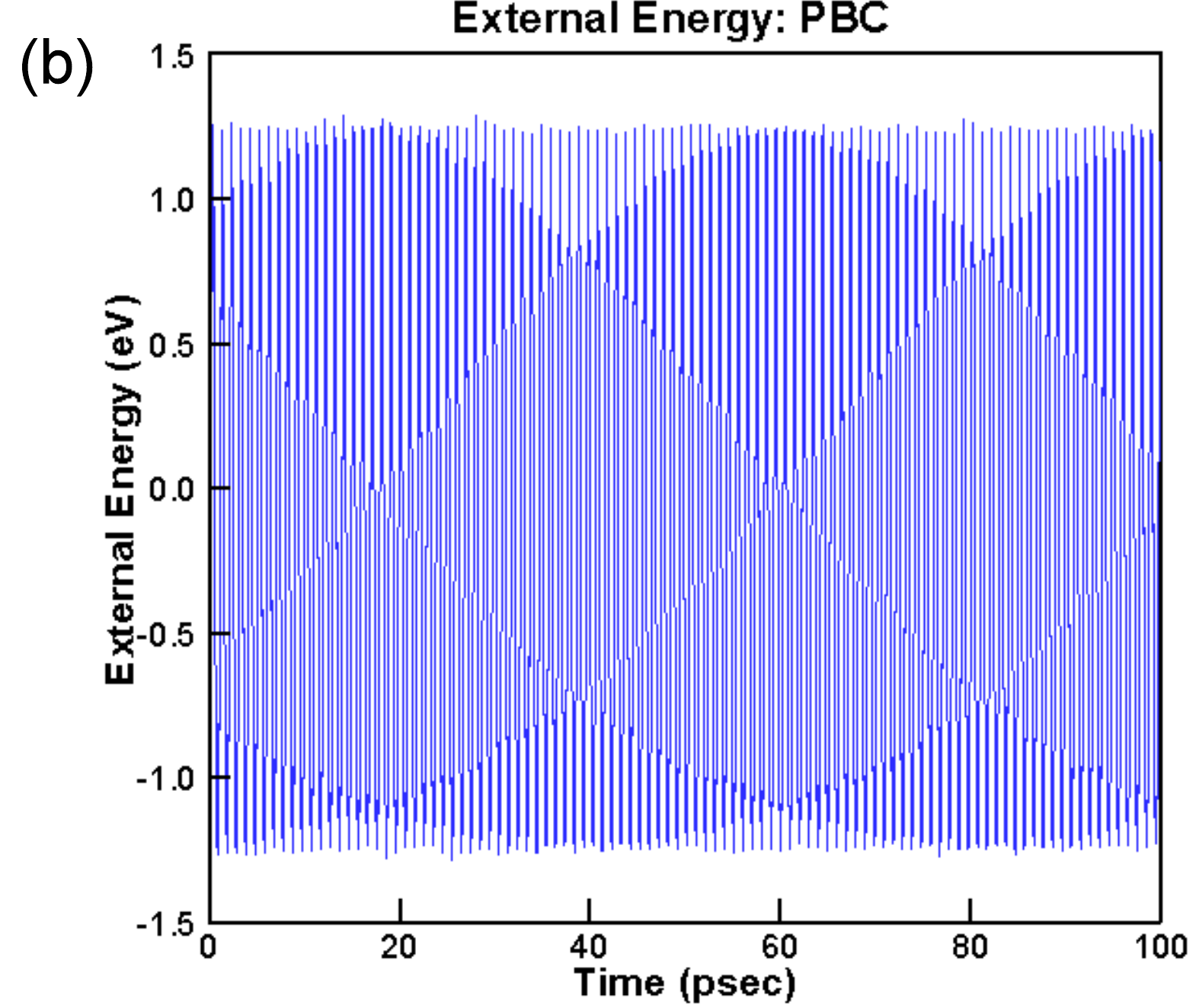}
\includegraphics[scale=0.54]{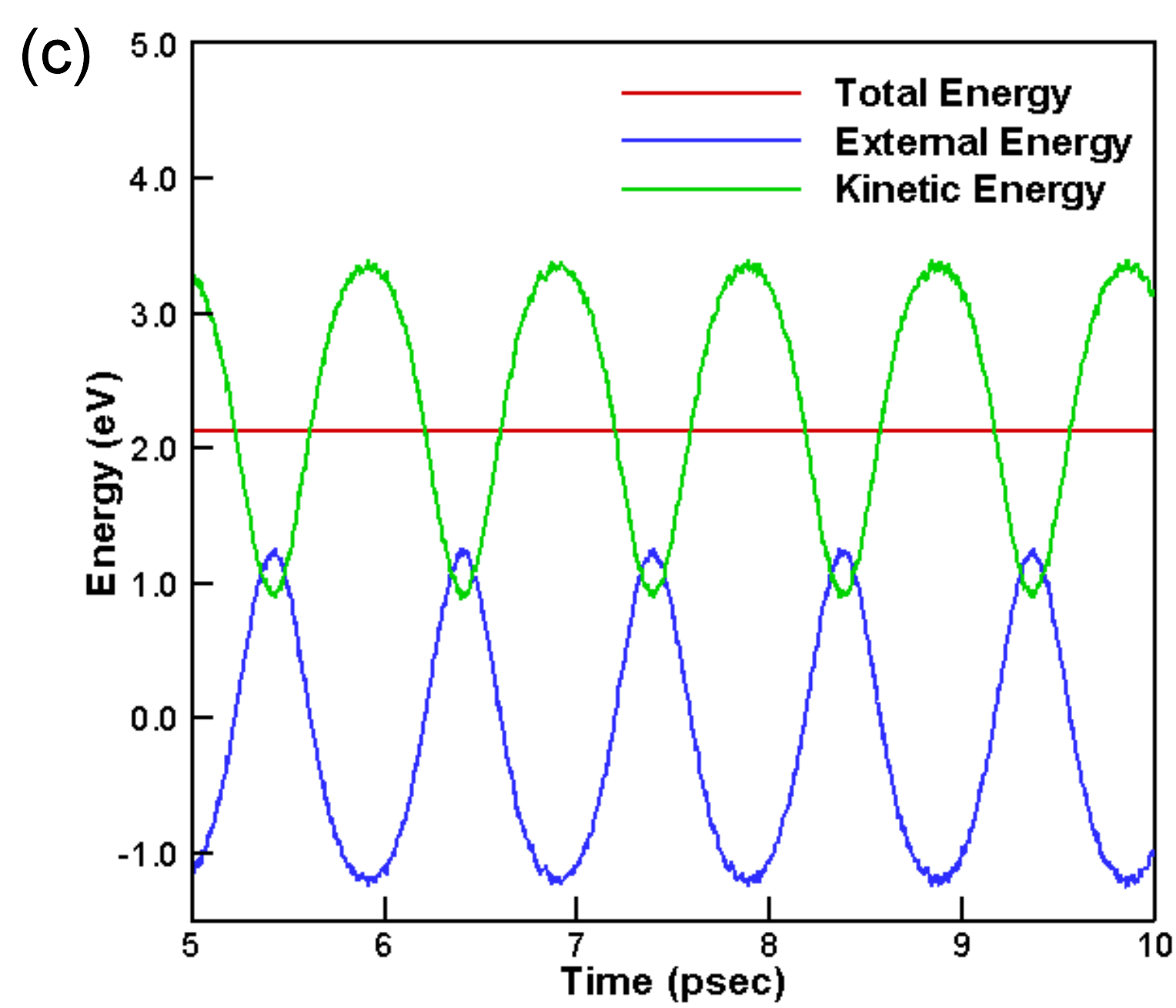}
\caption{\label{}(a) Snapshots of the coherent oscillation of a monolayer graphene sheet at 10 K with periodic boundary conditions applied.  (b) External energy time history showing minimal energy loss. (c) Close up of external energy, kinetic energy and total energy.}
\label{pbc} \end{center} \end{figure}

We utilized the second generation Brenner potential (REBO-II)~\cite{brennerJPCM2002} for both the carbon-carbon and carbon-hydrogen interactions; this potential has been shown to accurately reproduce binding energies, force constants and elastic properties of both graphene and hydrocarbons.  For all simulations, the graphene sheet was first equilibrated at a specified temperature using a Nose-Hoover thermostat~\cite{hooverPRA1985} for 50000 steps with a 1 femtosecond ($fs$) time step within an NVT ensemble.  After the initial thermal equilibration, a sinusoidal velocity profile was applied between the two long (fixed) edges of the graphene sheet, where the velocity was zero at the long (fixed) edges, and increased to a maximum value in the center of the graphene sheet.  After applying the sinusoidal velocity profile, the graphene monolayer was allowed to freely oscillate for 100000 steps in an energy conserving NVE ensemble.  The sinusoidal velocity that was applied to induce the oscillations was only 0.036\% of the total potential energy of the system, to ensure that nonlinear vibrational modes due to the applied velocity field would not be present.  

We first show the results at 10 K where PBCs were applied along the short edges of the graphene sheet.  As observed in Figure (\ref{pbc}a), the graphene sheet oscillates coherently, and completely in phase during the entire simulation time.  In particular, no difference in vibrational amplitude is observed between the edges and the center of the sheet at any time.  The coherency of oscillation and the lack of dissipation is confirmed by examining the external energy time history of the oscillation in Figure (\ref{pbc}b),  where the external energy is defined as the difference between the potential energy during the oscillation and the potential energy of the graphene sheet immediately after thermal equilibration and the application of the sinusoidal velocity field, but before the actual oscillation of the graphene monolayer has begun.  Therefore, oscillations of the external energy about the value of 0 eV, as in Figure (\ref{pbc}b), occur due to the corresponding oscillations of the graphene monolayer that occur due to the applied sinusoidal velocity field; any decrease in oscillation amplitude of the external energy, as we will show next for the free edge graphene monolayer, results from intrinsic loss mechanisms in the graphene monolayer.

As clearly shown in Figure (\ref{pbc}b), the coherency of the oscillation for the graphene monolayer with PBCs in Figure (\ref{pbc}a) results in a minimum of intrinsic energy dissipation.  Furthermore, as shown in Figure (\ref{pbc}c), the effects of the energy conserving NVE ensemble are observed, i.e. the total energy remains constant due to the out of phase oscillations exhibited by the kinetic and potential energies.

\begin{figure} \begin{center} 
\includegraphics[scale=0.5]{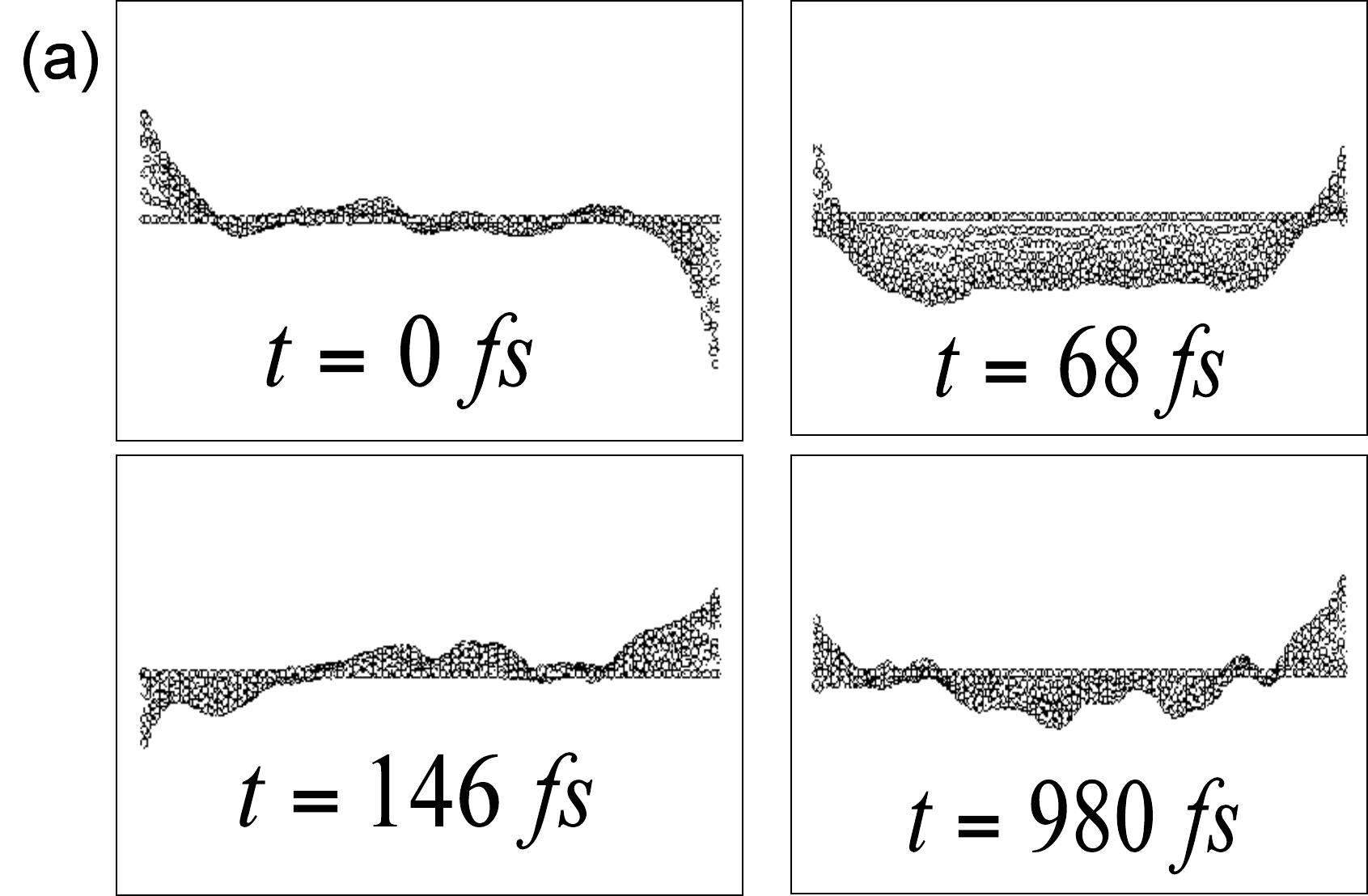}
\includegraphics[scale=0.54]{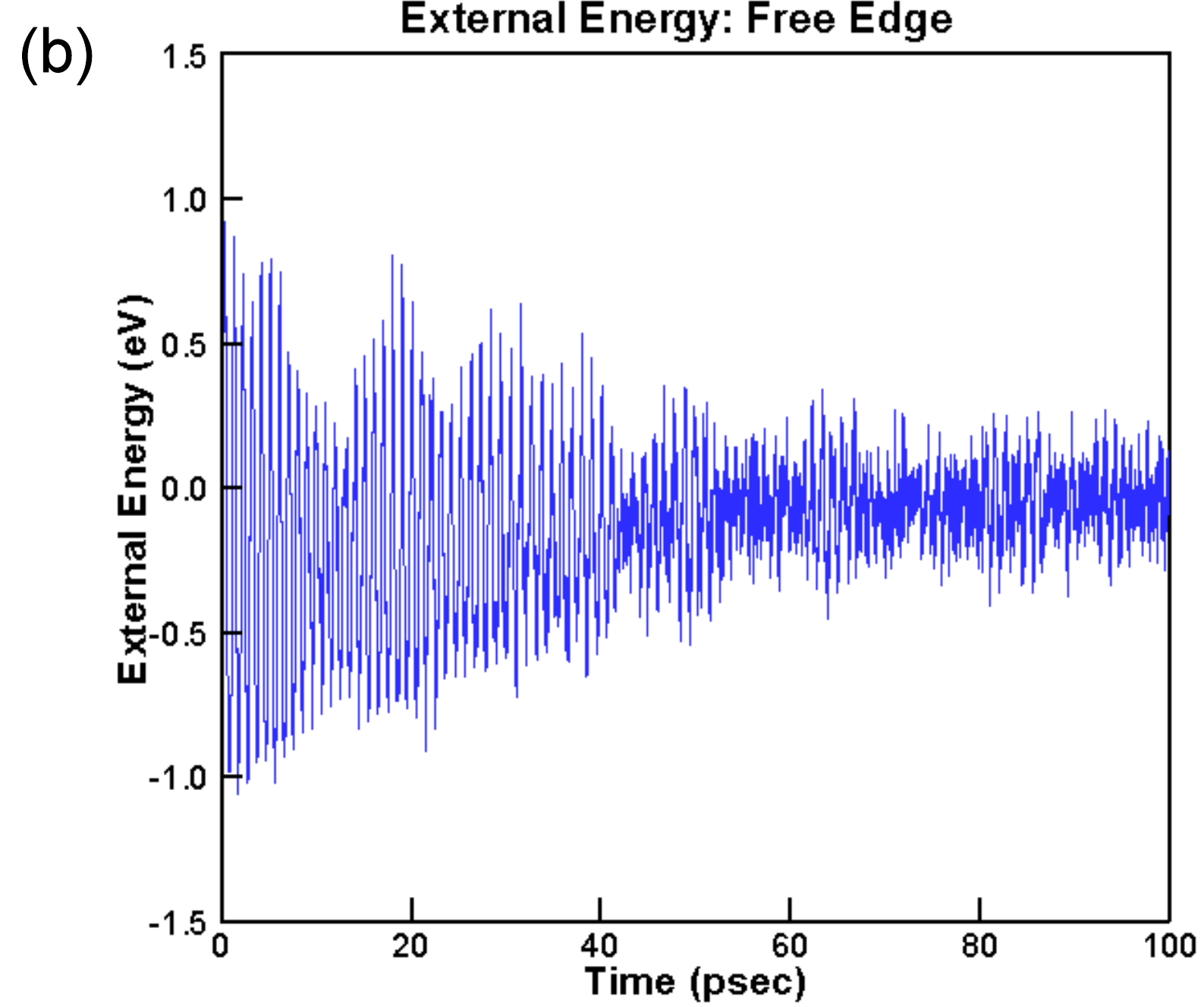}
\includegraphics[scale=0.54]{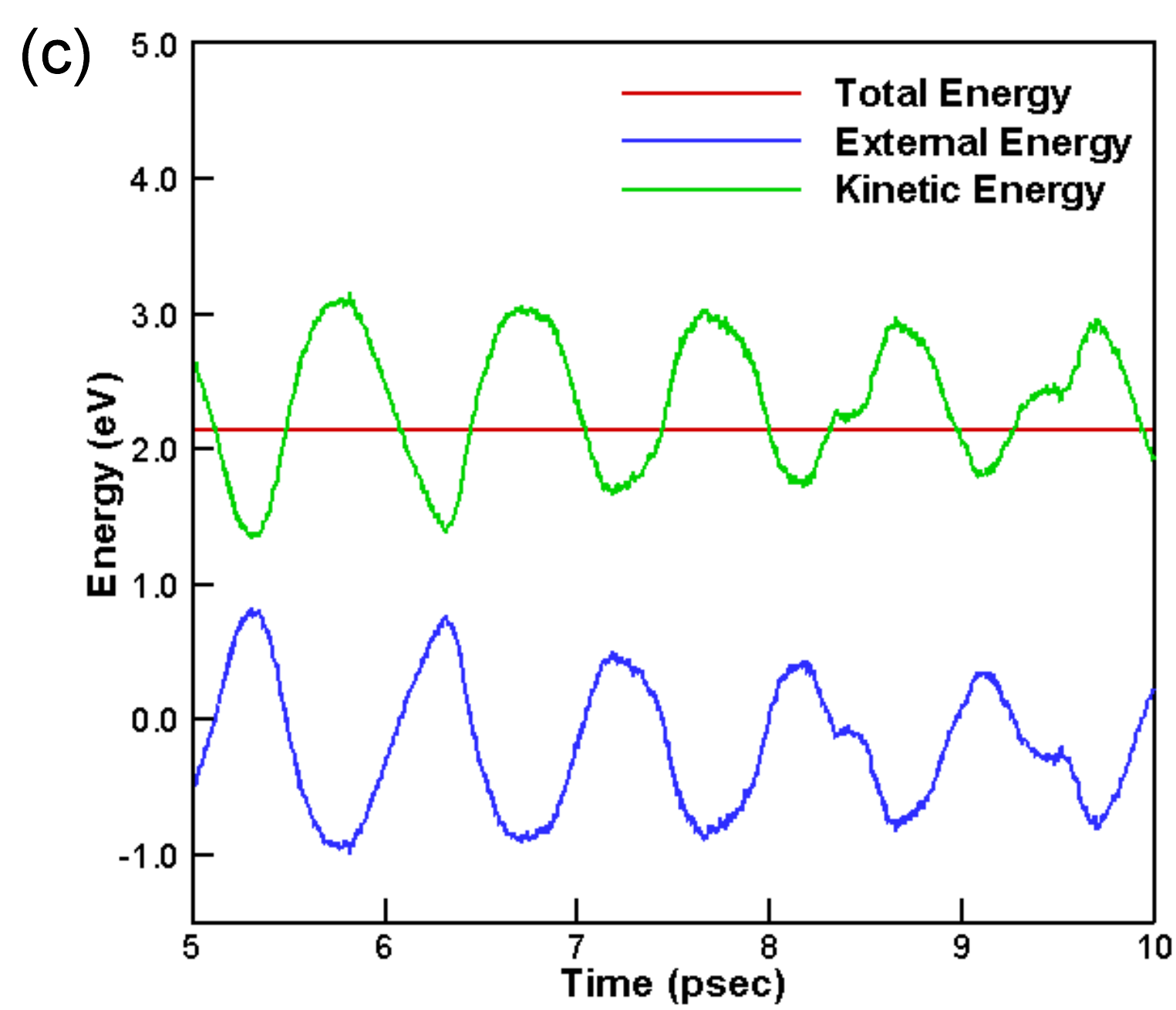}
\caption{\label{}(a) Snapshots of the incoherent oscillation of a monolayer graphene sheet at 10 K with two free edges showing mode mixing and beating phenomena.  (b) External energy time history showing significant energy dissipation and beating phenomena. (c) Close up of external energy, kinetic energy and total energy.}
\label{free_edge} \end{center} \end{figure}

We next show the results where the short edges of the graphene sheet were left free, or unconstrained; the edges had a zig-zag orientation.  This situation corresponds to the suspended graphene multilayers that have been tested experimentally by Bunch \emph{et al.}~\cite{bunchSCIENCE2007}, and Sanchez \emph{et al.}~\cite{sanchezNL2008}; suspension of the multilayer graphene sheet over a trench resulted in two edges of the graphene sheet being fixed (the fixing occurs experimentally through van der Waal's interactions with the underlying silicon oxide substrate), and two of the edges being free.  We note again the low Q-factors reported in both experimental studies.  

The results of the MD simulations for the oscillation of the graphene sheet at 10 K where the short edges are unrestrained are shown in Figure (\ref{free_edge}).  In comparing the results for the PBC case in Figure (\ref{pbc}a), it is clear that the oscillation time history for the graphene sheet with free edges in Figure (\ref{free_edge}a) is significantly different.  We first notice immediately, at $t=0$, that the free edges of the graphene sheet have a significantly larger vibrational amplitude than the interior of the sheet.  Furthermore, we notice at $t=0$ that the vibrational amplitudes of the free (short) edges are opposite in sign.  With an increase in simulation time, the spurious edge modes begin to propagate from the edges of the graphene sheet into the interior, thereby causing different portions of the graphene sheet to oscillate at different frequencies; the snapshots in Figure (\ref{free_edge}a) at $t=146 fs$ and $t=980 fs$ clearly illustrate how the edges, as well as different sections of the graphene sheet are oscillating incoherently and out of phase.  We note that during the MD simulations, the edges do not always have opposite vibrational amplitudes at $t=0$; due to the statistical nature of the Nose-Hoover thermostat~\cite{hooverPRA1985}, the initial vibrational amplitude at $t=0$ at the edges varies with each simulation.  However, what is constant in all simulations is that the spurious edge modes eventually dominate the oscillation of the graphene sheet, similar to that demonstrated for a specific case in Figure (\ref{free_edge}a).  

The effects of the incoherent oscillation induced by the free edge effects on the energy dissipation are shown in Figure (\ref{free_edge}b).  In comparing the external energy time history for the free edge case in Figure (\ref{free_edge}b) and the PBC case where there are no dangling bonds for the edge carbon atoms in Figure (\ref{pbc}b), we immediately see a significant increase in dissipation for the free edge case in Figure (\ref{free_edge}b).  Furthermore, we can detect a beating phenomenon for the free edge case in Figure (\ref{free_edge}b); the beating phenomenon manifests itself through the external energy whereby the energy does not monotonically decrease with increasing time, or the number of vibrational cycles.  Instead, the energy demonstrates peaks and valleys that do not correspond with the vibrational period of the graphene sheet; these peaks and valleys occur due to the incoherent mixing and interaction of disparate vibrational periods for the graphene sheet with free edges as seen in Figure (\ref{free_edge}a).  

Further insights can be obtained by comparing the kinetic, external and total energies in Figure (\ref{free_edge}c) as compared to the PBC energies in Figure (\ref{pbc}c).  We emphasize again that due to the NVE ensemble, the total energy is conserved in the free edge case, as shown in Figure (\ref{free_edge}c), just as it is in the PBC case in Figure (\ref{pbc}c).  However, the external energy and the kinetic energy are both observed to decrease dramatically in the free edge case in Figure (\ref{free_edge}c); the total energy remains constant while the components of the total energy are decreasing due to the out of phase oscillation of the kinetic and external energies.

We compare our results for the oscillation of the monolayer graphene sheet with free edges to that of Sanchez \emph{et al.}~\cite{sanchezNL2008}.  In that work, they also studied the dominance of edge oscillations.  However, in order to obtain the spurious edge modes, they externally applied a non-uniform stress and to the suspended graphene multilayer sheet in the form of both an in-plane stretch and an in-plane rotation.  Only after applying these states of stress were the spurious edge modes detected in their numerical (continuum finite element) simulations.

However, as shown in the present work, the spurious edge modes occur naturally through the MD simulations without application of any non-uniform stresses.  Instead, they arise due to the fact that the carbon atoms at the edges of the graphene sheet have fewer bonding neighbors, and are therefore undercoordinated with respect to the carbon atoms in the interior of the graphene sheet.  The lack of bonding neighbors means that the stiffness, and therefore the vibrational frequency of the edge atoms differs from the atoms within the graphene bulk; this difference in vibrational frequency of the edge atoms was illustrated in Figure (\ref{free_edge}a).  

We additionally note that both the free edge and PBC calculations were performed at 10 K, to minimize the effects of energy dissipation that are known to occur in oscillating nanoresonators with increasing temperature~\cite{evoyAPL2000}.  Because the energy dissipation, and thus Q-factor degradation occurs for the free edge graphene sheet at low temperature where thermal losses are known to be minimized, we have demonstrated that incoherent vibrational states resulting from edge effects are the dominant intrinsic loss mechanism for graphene nanoresonators.  

Because we have determined that the undercoordination of free edge carbon atoms in the graphene sheet is the key source for intrinsic damping effects in graphene nanoresonators, we performed simulations in which the edge carbon atoms were passivated with hydrogen, to remove the effects of undercoordination.  The validity of this idea is further enforced by recent works~\cite{junPRB2008,wassmannPRL2008}, both of which found that hydrogen passivation stabilizes the free edge graphene atoms, leaving the graphene edges essentially stress-free.  

\begin{figure} \begin{center} 
\includegraphics[scale=0.47]{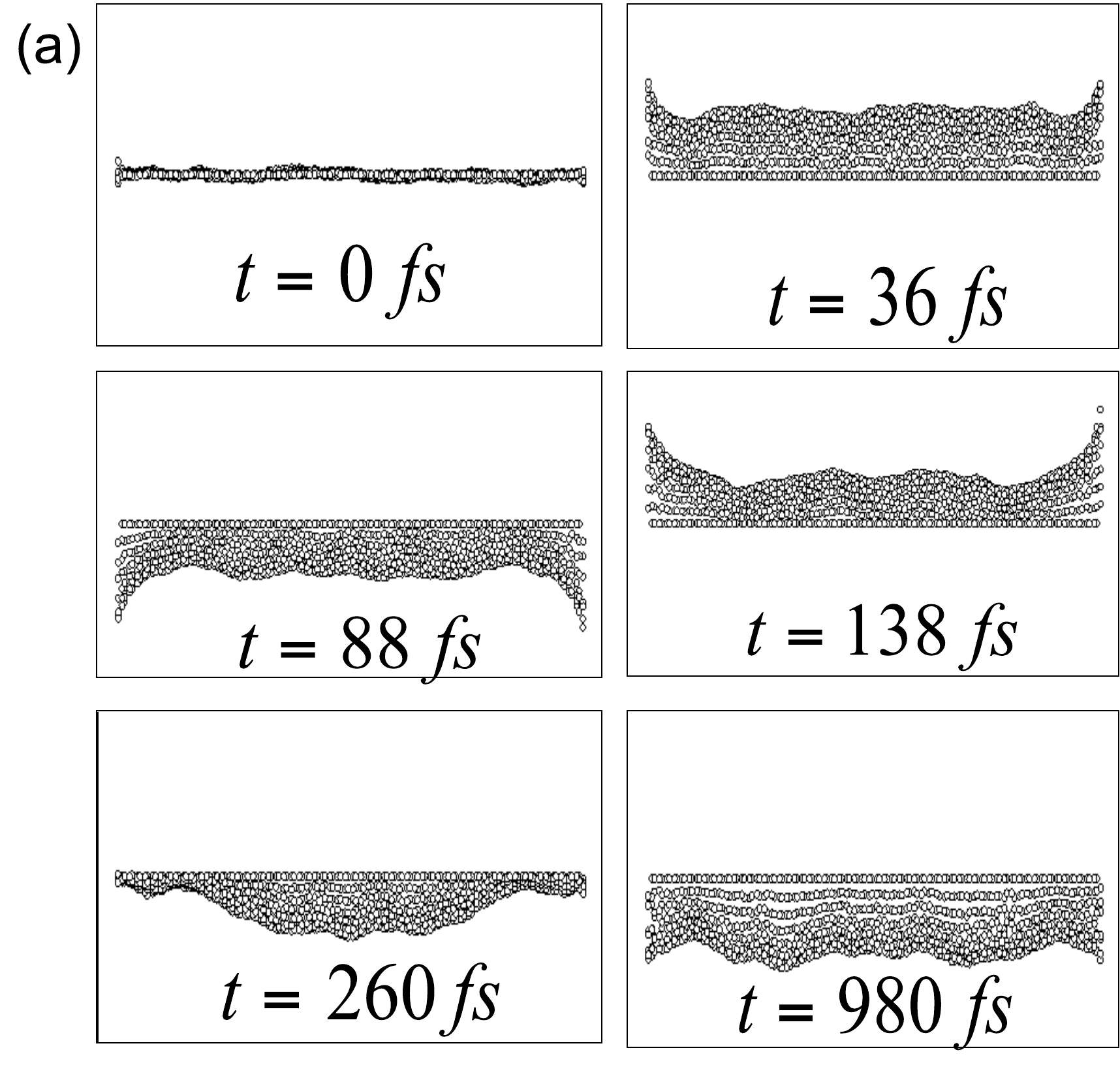}
\includegraphics[scale=0.54]{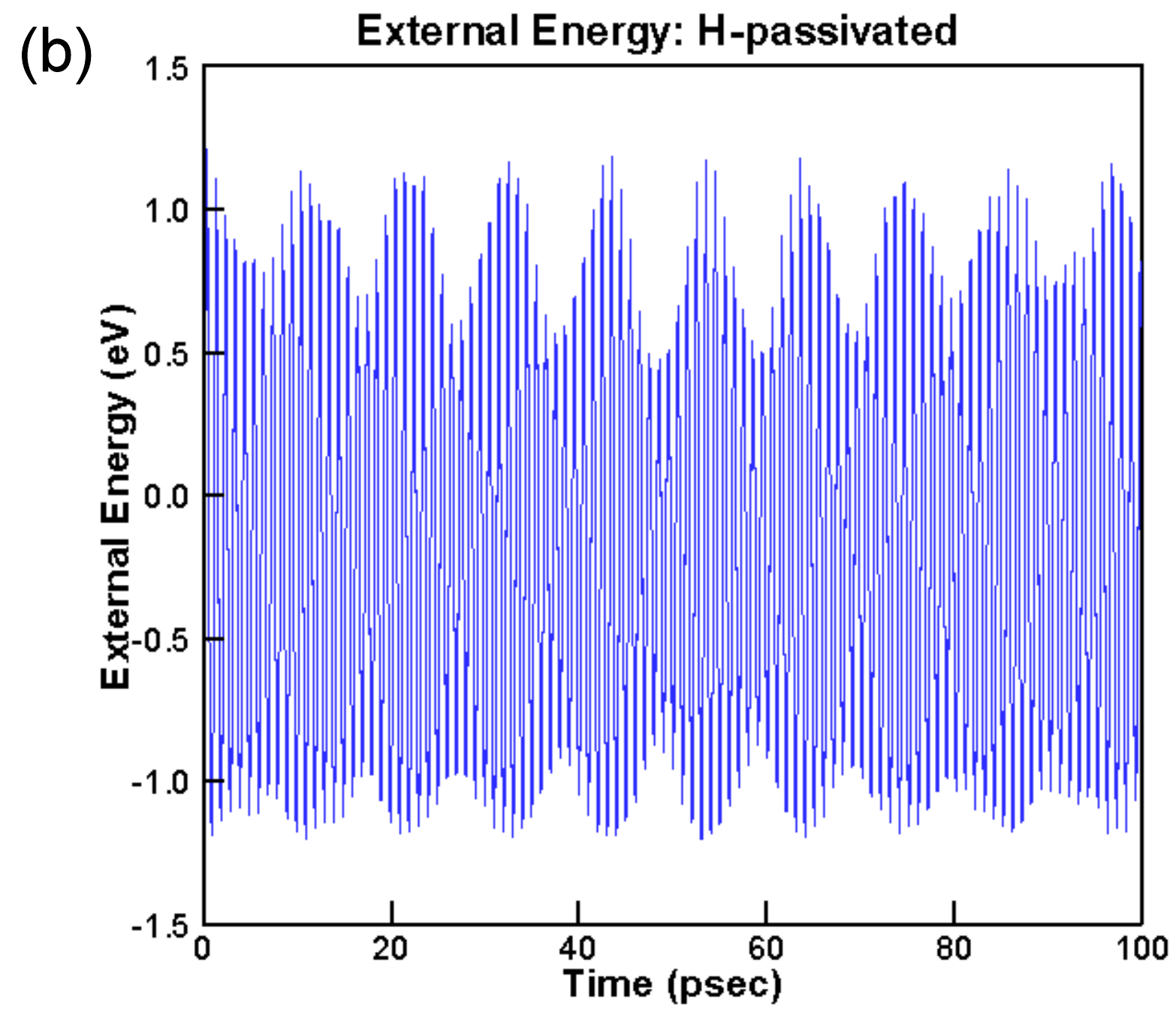}
\caption{\label{}(a) Snapshots of the oscillation of a monolayer graphene sheet at 10 K with its free edges passivated with hydrogen showing beating modes.  (b) External energy time history showing beating phenomena.}
\label{hpassivate} \end{center} \end{figure}

The results of the MD simulations of the oscillation of the graphene monolayer with hydrogen passivation at 10 K are shown in Figure (\ref{hpassivate}).  There are several noteworthy points in analyzing the oscillation history in Figure (\ref{hpassivate}a).  First, we notice that throughout the oscillation time history, the edges oscillate nearly in phase with the remainder of the graphene sheet.  However, the beating phenomena that was also observed for the free edge case in Figure (\ref{free_edge}b) is observed again in the oscillation time history and also the external energy time history in Figure (\ref{hpassivate}b).  The fact that different sections oscillate with different oscillation periods is most clearly observed at $t=980 fs$ in Figure (\ref{hpassivate}a), where the multiple minima and maxima of oscillation amplitude along the graphene sheet are observed.  Therefore, while hydrogen atoms can effectively energetically stabilize free edge carbon atoms that are undercoordinated, hydrogen passivation of dangling carbon bonds at the edges of the graphene sheet does not appear to be an optimal solution to mitigate free edge-induced energy dissipation in graphene nanoresonators.

Finally, we utilize MD simulations to study the intrinsic energy dissipation of monolayer graphene sheets where all edge atoms are constrained not to move.  In particular, we study circular monolayer graphene sheets that are similar geometrically to the circular multilayer graphene oxide sheets that have recently been fabricated and tested~\cite{robinsonNL2008a}, and where extremely high Q-factors with values up to 4000 have been found.  In addition, the graphene oxide multilayers studied by Robinson \emph{et al.}~\cite{robinsonNL2008a} were under tensile stress, which has recently proven beneficial in enhancing the Q-factors of both metallic~\cite{kimPRL2008} and semiconducting nanowires~\cite{verbridgeJAP2006,verbridgeNL2007,cimallaAPL2006}; however, it was not delineated whether the increase in Q found by Robinson \emph{et al.}~\cite{robinsonNL2008a} as compared to the suspended graphene multilayers tested earlier~\cite{bunchSCIENCE2007,sanchezNL2008} occurred due to the tensile stress, or due to the boundary conditions in which all edges of the multilayer graphene oxide sheet were fixed.  

We therefore studied the oscillations of a monolayer, circular graphene sheet with diameter of 42.6~\AA~that was comprised of 547 carbon atoms, in which all edge atoms were constrained not to move.  We imposed a sinusoidal velocity profile that was a maximum at the center of the sheet and decreased to zero at the circular edges to induce the required oscillatory motion, where again the kinetic energy of the applied sinusoidal wave was only 0.03\% of the total potential energy of the system to energy linear oscillatory motion.  In addition, we applied tensile strains of 1,2,3,4 and 5\% to the circular graphene sheet by radially expanding the sheet to determine the effects of tensile strain on the intrinsic energy dissipation of graphene.  

\begin{figure} \begin{center} 
\includegraphics[scale=0.54]{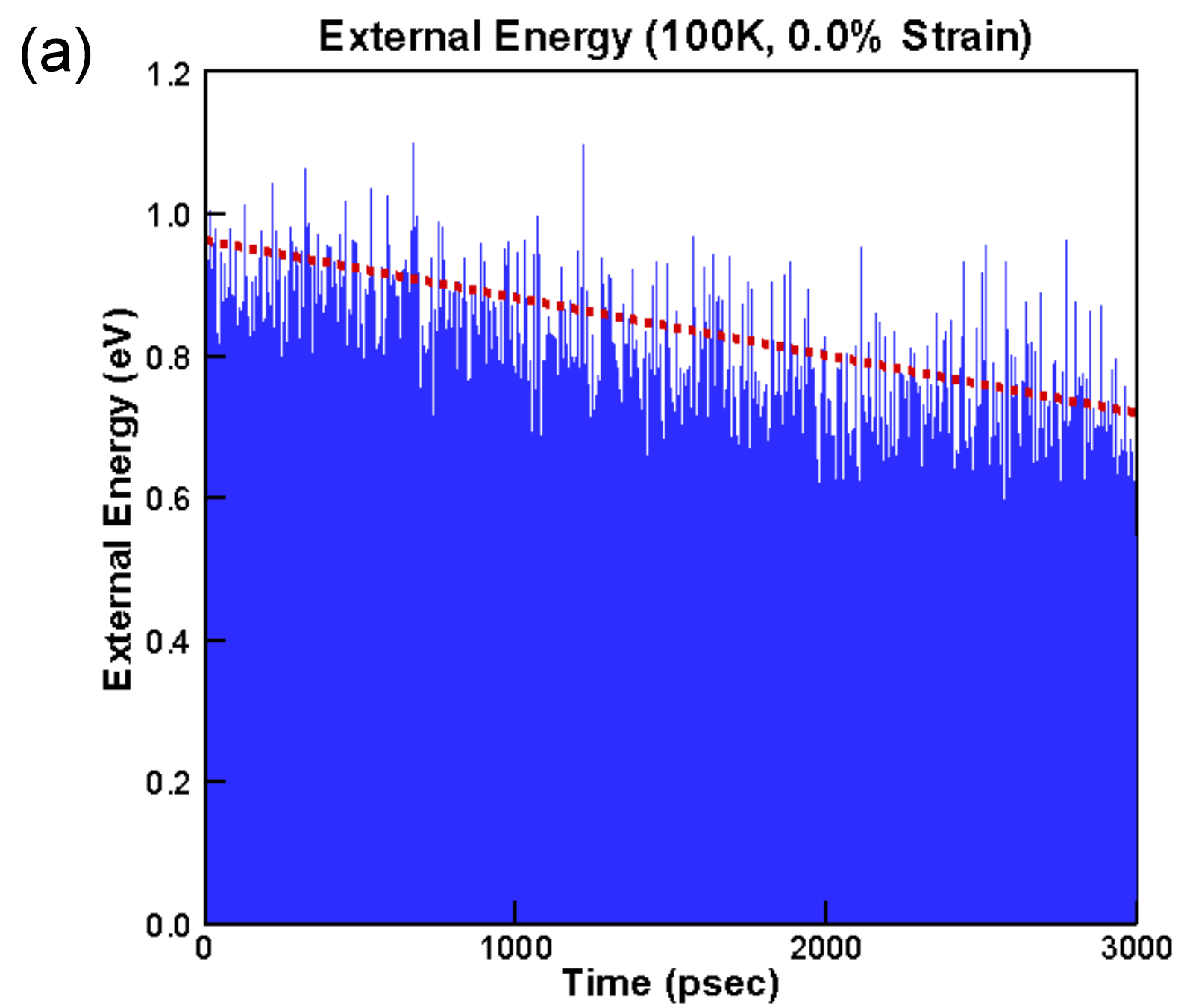}
\includegraphics[scale=0.54]{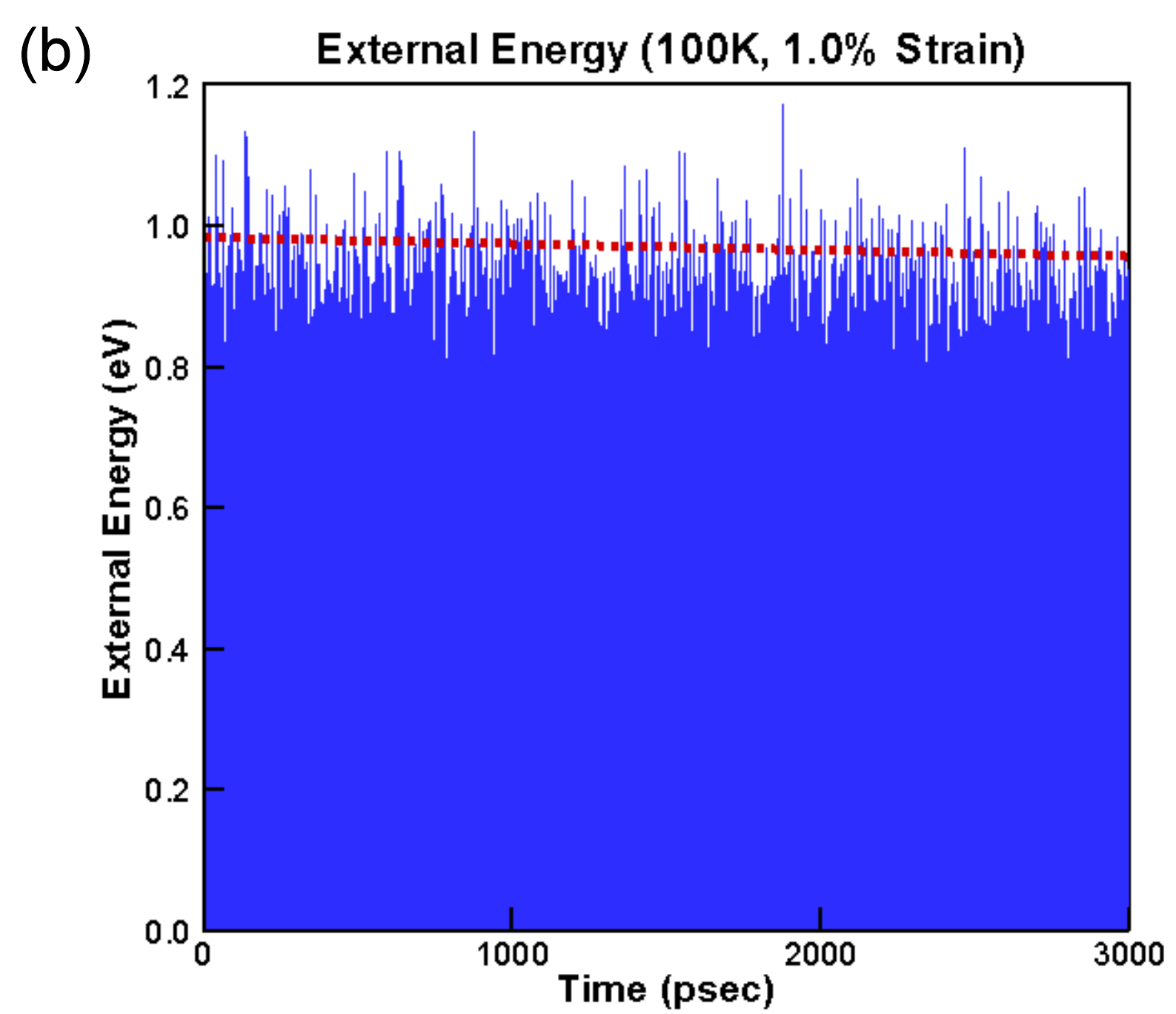}
\caption{\label{}External energy for circular graphene sheet at 100 K at (a) 0\% applied tensile strain, (b) 1\% applied tensile strain.  Dashed lines in both figures are drawn as a guide to the eye in quantifying the energy dissipation.}
\label{circlestrain} \end{center} \end{figure}

\begin{figure} \begin{center} 
\includegraphics[scale=0.41]{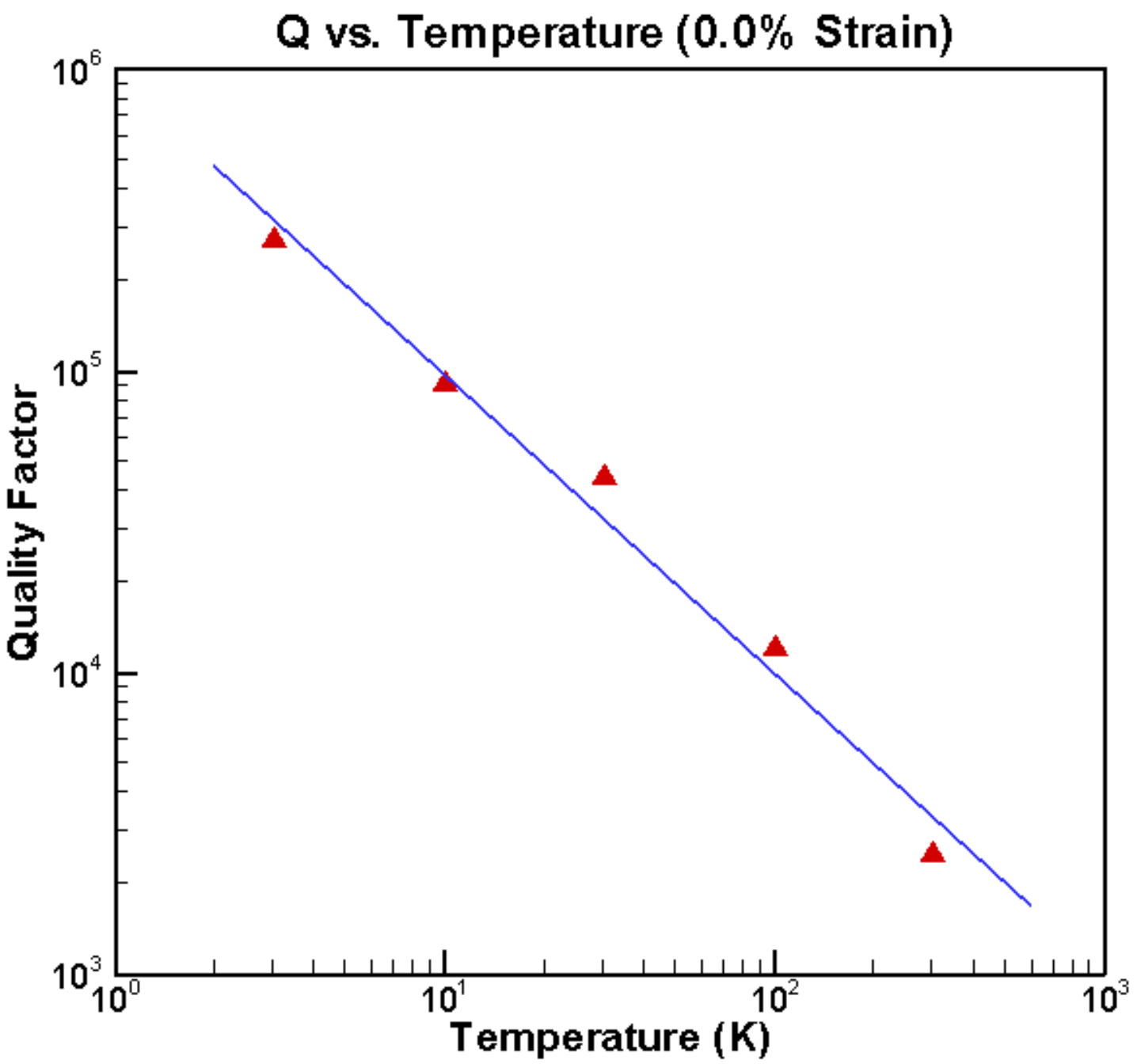}
\caption{\label{}Q-factor as a function of temperature for an unstrained circular graphene sheet.}
\label{circleq} \end{center} \end{figure}

The external energy time history for the circular graphene monolayer oscillating at 100 K and at 0\% and 1\% applied tensile strain is shown in Figure (\ref{circlestrain}).  As seen in Figure (\ref{circlestrain}a), when no tensile strain is applied, the graphene monolayer does dissipate energy, and the Q-factor is calculated to be about 12000.  However, we note that the energy dissipation is clearly smaller than the energy dissipation observed for the graphene sheet with free edges that was depicted in Figure (\ref{free_edge}b).  Furthermore, we note that the rapid energy dissipation in the free edge case in Figures (\ref{free_edge}b) was obtained after 100,000 time steps at a lower temperature (10 K); the present simulations were run for 3 million time steps in Figure (\ref{circlestrain}a) at 100 K, where significantly less energy dissipation is observed even at elevated temperature and over a time scale that is 30 times as long.  These results are clearly indicative of the fact that fixing the free edges of the graphene sheet, and thereby eliminating free edge vibrations, is the most effective way to remove spurious edge-induced intrinsic loss mechanisms in graphene nanoresonators.  

We show in Figure (\ref{circleq}) the variation in Q-factor for the circular graphene sheet as a function of temperature.  As can be seen, the Q-factor decreases with increasing temperature, i.e. from 273000 at 3K to about 2500 at 300 K.  We further find that the Q-factor is found to be inversely proportional to the temperature, i.e. $Q=1/T$.  The exponent of $1.0$ on the temperature $T$ differs from the exponent of $0.36$ found in previous MD simulations~\cite{jiangPRL2004} for the Q-factors of fixed/free carbon nanotubes.  The most likely explanation for the different damping exponent in the case of graphene is because we have removed all edge modes by fixing all edge atoms of the circular graphene sheet.  Because of this, and because therefore the remainder of the carbon atoms in the graphene sheet have a bulk-like bonding environment, the Q-factor degrades inversely proportional to the temperature as would be expected for a bulk material~\cite{mohantyPRB2002}.  In contrast, the fixed/free carbon nanotubes as studied by Jiang \emph{et al.}~\cite{jiangPRL2004} had a free end where the undercoordinated carbon atoms were not constrained; the non-bulk bonding environment of the free edge atoms of the nanotube are the likely cause of the non-bulk temperature damping exponent as compared to a bulk material.

We also demonstrate in Figures (\ref{circlestrain}b) the external energy time history for the circular graphene sheet at 1\% tensile strain.  As can be observed, there is very minimal energy loss when at 1\% tensile strain; applying larger amounts of tensile also leads to very small intrinsic energy losses during the oscillation of the circular graphene sheet.  This phenomenon was also observed for other
geometries (square graphene sheets), and a wide range of temperatures from 3-300 K.  This result for graphene is in line with recent theoretical results on metal nanowires~\cite{kimPRL2008}, and experimental results where the Q-factors of SiN, Si and SiC nanowires have been elevated nearly an order of magnitude through the application of tensile stress~\cite{verbridgeJAP2006,verbridgeNL2007,cimallaAPL2006}, though the mechanisms causing the increase in Q are different between graphene and the nanowires.  

Experimentally, it was argued by Verbridge \emph{et al.}~\cite{verbridgeNL2007} that adding tensile stress to SiN nanowires reduced clamping losses by increasing acoustic mismatch between the resonating nanowire and the supporting substrate.  In the present work, the increase in Q-factor likely results from the fact that increased tensile strain in the graphene sheets mitigates thermally-driven fluctuations and local variations in vibrational frequency.  Therefore, the present simulations show that tensile strain, in combination of the removal of dangling bonds for carbon atoms that lie on the edges of the graphene sheet, can be utilized to essentially remove all intrinsic loss mechanisms in graphene nanoresonators.  

\section{Conclusions}

In conclusion, we have utilized classical molecular dynamics to study the intrinsic loss mechanisms, and therefore the reasons underlying the Q-factor degradation in monolayer graphene nanoresonators undergoing flexural oscillations.  In doing so, we have determined that spurious edge modes of vibration, which arise not due to externally applied stresses but intrinsically due to the different vibrational properties of edge atoms, are the dominant intrinsic loss mechanism that reduces the Q-factors.  We further determined that hydrogen passivation is ineffective due to the persistance of the beating phenomena that is also observed for the free edge, or suspended graphene sheets.  Finally, we determined that in the absence of applied tensile strain, the Q-factor degrades inversely proportional to temperature; however, it was determined that the intrinsic loss mechanisms in graphene can nearly be eliminated through a combination of applied tensile mechanical strain and boundary conditions, i.e. fixing all edge atoms. 

\section*{Acknowledgements}

SYK and HSP both gratefully acknowledge the support of DARPA through grant HR0011-08-1-0047.  The authors note that the content of this article does not necessarily reflect the position or the policy of the government, and that no official government endorsement of the results should be inferred.  HSP also acknowledges support from NSF grant CMMI-0750395.  Both authors acknowledge helpful discussions with J. Scott Bunch, and Prof. Sulin Zhang for sharing his Brenner code.


\begin{thebibliography}{23}
\expandafter\ifx\csname natexlab\endcsname\relax\def\natexlab#1{#1}\fi
\expandafter\ifx\csname bibnamefont\endcsname\relax
  \def\bibnamefont#1{#1}\fi
\expandafter\ifx\csname bibfnamefont\endcsname\relax
  \def\bibfnamefont#1{#1}\fi
\expandafter\ifx\csname citenamefont\endcsname\relax
  \def\citenamefont#1{#1}\fi
\expandafter\ifx\csname url\endcsname\relax
  \def\url#1{\texttt{#1}}\fi
\expandafter\ifx\csname urlprefix\endcsname\relax\def\urlprefix{URL }\fi
\providecommand{\bibinfo}[2]{#2}
\providecommand{\eprint}[2][]{\url{#2}}

\bibitem[{\citenamefont{Novoselov
  et~al.}(2005{\natexlab{a}})\citenamefont{Novoselov, Geim, Morozov, Jiang,
  Katsnelson, Grigorieva, Dubonos, and Firsov}}]{novoselovNATURE2005}
\bibinfo{author}{\bibfnamefont{K.~S.} \bibnamefont{Novoselov}},
  \bibinfo{author}{\bibfnamefont{A.~K.} \bibnamefont{Geim}},
  \bibinfo{author}{\bibfnamefont{S.~V.} \bibnamefont{Morozov}},
  \bibinfo{author}{\bibfnamefont{D.}~\bibnamefont{Jiang}},
  \bibinfo{author}{\bibfnamefont{M.~I.} \bibnamefont{Katsnelson}},
  \bibinfo{author}{\bibfnamefont{I.~V.} \bibnamefont{Grigorieva}},
  \bibinfo{author}{\bibfnamefont{S.~V.} \bibnamefont{Dubonos}},
  \bibnamefont{and} \bibinfo{author}{\bibfnamefont{A.~A.}
  \bibnamefont{Firsov}}, \bibinfo{journal}{Nature}
  \textbf{\bibinfo{volume}{438}}, \bibinfo{pages}{197}
  (\bibinfo{year}{2005}{\natexlab{a}}).

\bibitem[{\citenamefont{Novoselov
  et~al.}(2005{\natexlab{b}})\citenamefont{Novoselov, Jiang, Schedin, Booth,
  Khotkevich, Morozov, and Geim}}]{novoselovPNAS2005}
\bibinfo{author}{\bibfnamefont{K.~S.} \bibnamefont{Novoselov}},
  \bibinfo{author}{\bibfnamefont{D.}~\bibnamefont{Jiang}},
  \bibinfo{author}{\bibfnamefont{F.}~\bibnamefont{Schedin}},
  \bibinfo{author}{\bibfnamefont{T.~J.} \bibnamefont{Booth}},
  \bibinfo{author}{\bibfnamefont{V.~V.} \bibnamefont{Khotkevich}},
  \bibinfo{author}{\bibfnamefont{S.~V.} \bibnamefont{Morozov}},
  \bibnamefont{and} \bibinfo{author}{\bibfnamefont{A.~K.} \bibnamefont{Geim}},
  \bibinfo{journal}{Proceedings of the National Academy of Science}
  \textbf{\bibinfo{volume}{102}}, \bibinfo{pages}{10451}
  (\bibinfo{year}{2005}{\natexlab{b}}).

\bibitem[{\citenamefont{Geim and Novoselov}(2007)}]{geimNM2007}
\bibinfo{author}{\bibfnamefont{A.~K.} \bibnamefont{Geim}} \bibnamefont{and}
  \bibinfo{author}{\bibfnamefont{K.~S.} \bibnamefont{Novoselov}},
  \bibinfo{journal}{Nature Materials} \textbf{\bibinfo{volume}{6}},
  \bibinfo{pages}{183} (\bibinfo{year}{2007}).

\bibitem[{\citenamefont{Stankovich et~al.}(2006)\citenamefont{Stankovich,
  Dikin, Dommett, Kohlhaas, Zimney, Stach, Piner, Nguyen, and
  Ruoff}}]{stankovichNATURE2006}
\bibinfo{author}{\bibfnamefont{S.}~\bibnamefont{Stankovich}},
  \bibinfo{author}{\bibfnamefont{D.~A.} \bibnamefont{Dikin}},
  \bibinfo{author}{\bibfnamefont{G.~H.~B.} \bibnamefont{Dommett}},
  \bibinfo{author}{\bibfnamefont{K.~M.} \bibnamefont{Kohlhaas}},
  \bibinfo{author}{\bibfnamefont{E.~J.} \bibnamefont{Zimney}},
  \bibinfo{author}{\bibfnamefont{E.~A.} \bibnamefont{Stach}},
  \bibinfo{author}{\bibfnamefont{R.~D.} \bibnamefont{Piner}},
  \bibinfo{author}{\bibfnamefont{S.~T.} \bibnamefont{Nguyen}},
  \bibnamefont{and} \bibinfo{author}{\bibfnamefont{R.~S.} \bibnamefont{Ruoff}},
  \bibinfo{journal}{Nature} \textbf{\bibinfo{volume}{442}},
  \bibinfo{pages}{282} (\bibinfo{year}{2006}).

\bibitem[{\citenamefont{Bunch et~al.}(2007)\citenamefont{Bunch, van~der Zande,
  Verbridge, Frank, Tanenbaum, Parpia, Craighead, and
  McEuen}}]{bunchSCIENCE2007}
\bibinfo{author}{\bibfnamefont{J.~S.} \bibnamefont{Bunch}},
  \bibinfo{author}{\bibfnamefont{A.~M.} \bibnamefont{van~der Zande}},
  \bibinfo{author}{\bibfnamefont{S.~S.} \bibnamefont{Verbridge}},
  \bibinfo{author}{\bibfnamefont{I.~W.} \bibnamefont{Frank}},
  \bibinfo{author}{\bibfnamefont{D.~M.} \bibnamefont{Tanenbaum}},
  \bibinfo{author}{\bibfnamefont{J.~M.} \bibnamefont{Parpia}},
  \bibinfo{author}{\bibfnamefont{H.~G.} \bibnamefont{Craighead}},
  \bibnamefont{and} \bibinfo{author}{\bibfnamefont{P.~L.}
  \bibnamefont{McEuen}}, \bibinfo{journal}{Science}
  \textbf{\bibinfo{volume}{315}}, \bibinfo{pages}{490} (\bibinfo{year}{2007}).

\bibitem[{\citenamefont{Frank et~al.}(2007)\citenamefont{Frank, Tanenbaum,
  van~der Zande, and McEuen}}]{frankJVSTB2007}
\bibinfo{author}{\bibfnamefont{I.~W.} \bibnamefont{Frank}},
  \bibinfo{author}{\bibfnamefont{D.~M.} \bibnamefont{Tanenbaum}},
  \bibinfo{author}{\bibfnamefont{A.~M.} \bibnamefont{van~der Zande}},
  \bibnamefont{and} \bibinfo{author}{\bibfnamefont{P.~L.}
  \bibnamefont{McEuen}}, \bibinfo{journal}{Journal of Vacuum Science and
  Technology B} \textbf{\bibinfo{volume}{25}}, \bibinfo{pages}{2558}
  (\bibinfo{year}{2007}).

\bibitem[{\citenamefont{Garcia-Sanchez
  et~al.}(2008)\citenamefont{Garcia-Sanchez, van~der Zande, Paulo, Lassagne,
  McEuen, and Bachtold}}]{sanchezNL2008}
\bibinfo{author}{\bibfnamefont{D.}~\bibnamefont{Garcia-Sanchez}},
  \bibinfo{author}{\bibfnamefont{A.~M.} \bibnamefont{van~der Zande}},
  \bibinfo{author}{\bibfnamefont{A.~S.} \bibnamefont{Paulo}},
  \bibinfo{author}{\bibfnamefont{B.}~\bibnamefont{Lassagne}},
  \bibinfo{author}{\bibfnamefont{P.~L.} \bibnamefont{McEuen}},
  \bibnamefont{and} \bibinfo{author}{\bibfnamefont{A.}~\bibnamefont{Bachtold}},
  \bibinfo{journal}{Nano Letters} \textbf{\bibinfo{volume}{8}},
  \bibinfo{pages}{1399} (\bibinfo{year}{2008}).

\bibitem[{\citenamefont{Robinson et~al.}(2008)\citenamefont{Robinson,
  Zalalutdinov, Baldwin, Snow, Wei, Sheehan, and Houston}}]{robinsonNL2008a}
\bibinfo{author}{\bibfnamefont{J.~T.} \bibnamefont{Robinson}},
  \bibinfo{author}{\bibfnamefont{M.}~\bibnamefont{Zalalutdinov}},
  \bibinfo{author}{\bibfnamefont{J.~W.} \bibnamefont{Baldwin}},
  \bibinfo{author}{\bibfnamefont{E.~S.} \bibnamefont{Snow}},
  \bibinfo{author}{\bibfnamefont{Z.}~\bibnamefont{Wei}},
  \bibinfo{author}{\bibfnamefont{P.}~\bibnamefont{Sheehan}}, \bibnamefont{and}
  \bibinfo{author}{\bibfnamefont{B.~H.} \bibnamefont{Houston}},
  \bibinfo{journal}{Nano Letters} \textbf{\bibinfo{volume}{8}},
  \bibinfo{pages}{3441} (\bibinfo{year}{2008}).

\bibitem[{\citenamefont{Lee et~al.}(2008)\citenamefont{Lee, Wei, Kysar, and
  Hone}}]{leeSCIENCE2008}
\bibinfo{author}{\bibfnamefont{C.}~\bibnamefont{Lee}},
  \bibinfo{author}{\bibfnamefont{X.}~\bibnamefont{Wei}},
  \bibinfo{author}{\bibfnamefont{J.~W.} \bibnamefont{Kysar}}, \bibnamefont{and}
  \bibinfo{author}{\bibfnamefont{J.}~\bibnamefont{Hone}},
  \bibinfo{journal}{Science} \textbf{\bibinfo{volume}{321}},
  \bibinfo{pages}{385} (\bibinfo{year}{2008}).

\bibitem[{\citenamefont{Stowe et~al.}(1997)\citenamefont{Stowe, Yasumura,
  Kenny, Botkin, Wago, and Rugar}}]{stoweAPL1997}
\bibinfo{author}{\bibfnamefont{T.~D.} \bibnamefont{Stowe}},
  \bibinfo{author}{\bibfnamefont{K.}~\bibnamefont{Yasumura}},
  \bibinfo{author}{\bibfnamefont{T.~W.} \bibnamefont{Kenny}},
  \bibinfo{author}{\bibfnamefont{D.}~\bibnamefont{Botkin}},
  \bibinfo{author}{\bibfnamefont{K.}~\bibnamefont{Wago}}, \bibnamefont{and}
  \bibinfo{author}{\bibfnamefont{D.}~\bibnamefont{Rugar}},
  \bibinfo{journal}{Applied Physics Letters} \textbf{\bibinfo{volume}{71}},
  \bibinfo{pages}{288} (\bibinfo{year}{1997}).

\bibitem[{\citenamefont{Ekinci and Roukes}(2005)}]{ekinciRSI2005}
\bibinfo{author}{\bibfnamefont{K.~L.} \bibnamefont{Ekinci}} \bibnamefont{and}
  \bibinfo{author}{\bibfnamefont{M.~L.} \bibnamefont{Roukes}},
  \bibinfo{journal}{Review of Scientific Instruments}
  \textbf{\bibinfo{volume}{76}}, \bibinfo{pages}{061101}
  (\bibinfo{year}{2005}).

\bibitem[{\citenamefont{Brenner et~al.}(2002)\citenamefont{Brenner, Shenderova,
  Harrison, Stuart, Ni, and Sinnott}}]{brennerJPCM2002}
\bibinfo{author}{\bibfnamefont{D.~W.} \bibnamefont{Brenner}},
  \bibinfo{author}{\bibfnamefont{O.~A.} \bibnamefont{Shenderova}},
  \bibinfo{author}{\bibfnamefont{J.~A.} \bibnamefont{Harrison}},
  \bibinfo{author}{\bibfnamefont{S.~J.} \bibnamefont{Stuart}},
  \bibinfo{author}{\bibfnamefont{B.}~\bibnamefont{Ni}}, \bibnamefont{and}
  \bibinfo{author}{\bibfnamefont{S.~B.} \bibnamefont{Sinnott}},
  \bibinfo{journal}{Journal of Physics: Condensed Matter}
  \textbf{\bibinfo{volume}{14}}, \bibinfo{pages}{783} (\bibinfo{year}{2002}).

\bibitem[{\citenamefont{Yasumura et~al.}(2000)\citenamefont{Yasumura, Stowe,
  Chow, Pfafman, Kenny, Stipe, and Rugar}}]{yasumuraJMS2000}
\bibinfo{author}{\bibfnamefont{K.~Y.} \bibnamefont{Yasumura}},
  \bibinfo{author}{\bibfnamefont{T.~D.} \bibnamefont{Stowe}},
  \bibinfo{author}{\bibfnamefont{E.~M.} \bibnamefont{Chow}},
  \bibinfo{author}{\bibfnamefont{T.}~\bibnamefont{Pfafman}},
  \bibinfo{author}{\bibfnamefont{T.~W.} \bibnamefont{Kenny}},
  \bibinfo{author}{\bibfnamefont{B.~C.} \bibnamefont{Stipe}}, \bibnamefont{and}
  \bibinfo{author}{\bibfnamefont{D.}~\bibnamefont{Rugar}},
  \bibinfo{journal}{Journal of Microelectromechanical Systems}
  \textbf{\bibinfo{volume}{9}}, \bibinfo{pages}{117} (\bibinfo{year}{2000}).

\bibitem[{\citenamefont{Hoover}(1985)}]{hooverPRA1985}
\bibinfo{author}{\bibfnamefont{W.~G.} \bibnamefont{Hoover}},
  \bibinfo{journal}{Physical Review A} \textbf{\bibinfo{volume}{31}},
  \bibinfo{pages}{1695} (\bibinfo{year}{1985}).

\bibitem[{\citenamefont{Evoy et~al.}(2000)\citenamefont{Evoy, Olkhovets,
  Sekaric, Parpia, Craighead, and Carr}}]{evoyAPL2000}
\bibinfo{author}{\bibfnamefont{S.}~\bibnamefont{Evoy}},
  \bibinfo{author}{\bibfnamefont{A.}~\bibnamefont{Olkhovets}},
  \bibinfo{author}{\bibfnamefont{L.}~\bibnamefont{Sekaric}},
  \bibinfo{author}{\bibfnamefont{J.~M.} \bibnamefont{Parpia}},
  \bibinfo{author}{\bibfnamefont{H.~G.} \bibnamefont{Craighead}},
  \bibnamefont{and} \bibinfo{author}{\bibfnamefont{D.~W.} \bibnamefont{Carr}},
  \bibinfo{journal}{Applied Physics Letters} \textbf{\bibinfo{volume}{77}},
  \bibinfo{pages}{2397} (\bibinfo{year}{2000}).

\bibitem[{\citenamefont{Jun}(2008)}]{junPRB2008}
\bibinfo{author}{\bibfnamefont{S.}~\bibnamefont{Jun}},
  \bibinfo{journal}{Physical Review B} \textbf{\bibinfo{volume}{78}},
  \bibinfo{pages}{073405} (\bibinfo{year}{2008}).

\bibitem[{\citenamefont{Wassmann et~al.}(2008)\citenamefont{Wassmann,
  Seitsonen, Saitta, Lazzeri, and Mauri}}]{wassmannPRL2008}
\bibinfo{author}{\bibfnamefont{T.}~\bibnamefont{Wassmann}},
  \bibinfo{author}{\bibfnamefont{A.~P.} \bibnamefont{Seitsonen}},
  \bibinfo{author}{\bibfnamefont{A.~M.} \bibnamefont{Saitta}},
  \bibinfo{author}{\bibfnamefont{M.}~\bibnamefont{Lazzeri}}, \bibnamefont{and}
  \bibinfo{author}{\bibfnamefont{F.}~\bibnamefont{Mauri}},
  \bibinfo{journal}{Physical Review Letters} \textbf{\bibinfo{volume}{101}},
  \bibinfo{pages}{096402} (\bibinfo{year}{2008}).

\bibitem[{\citenamefont{Kim and Park}(2008)}]{kimPRL2008}
\bibinfo{author}{\bibfnamefont{S.~Y.} \bibnamefont{Kim}} \bibnamefont{and}
  \bibinfo{author}{\bibfnamefont{H.~S.} \bibnamefont{Park}},
  \bibinfo{journal}{Physical Review Letters} \textbf{\bibinfo{volume}{101}},
  \bibinfo{pages}{215502} (\bibinfo{year}{2008}).

\bibitem[{\citenamefont{Verbridge et~al.}(2006)\citenamefont{Verbridge, Parpia,
  Reichenbach, Bellan, and Craighead}}]{verbridgeJAP2006}
\bibinfo{author}{\bibfnamefont{S.~S.} \bibnamefont{Verbridge}},
  \bibinfo{author}{\bibfnamefont{J.~M.} \bibnamefont{Parpia}},
  \bibinfo{author}{\bibfnamefont{R.~B.} \bibnamefont{Reichenbach}},
  \bibinfo{author}{\bibfnamefont{L.~M.} \bibnamefont{Bellan}},
  \bibnamefont{and} \bibinfo{author}{\bibfnamefont{H.~G.}
  \bibnamefont{Craighead}}, \bibinfo{journal}{Journal of Applied Physics}
  \textbf{\bibinfo{volume}{99}}, \bibinfo{pages}{124304}
  (\bibinfo{year}{2006}).

\bibitem[{\citenamefont{Verbridge et~al.}(2007)\citenamefont{Verbridge,
  Shapiro, Craighead, and Parpia}}]{verbridgeNL2007}
\bibinfo{author}{\bibfnamefont{S.~S.} \bibnamefont{Verbridge}},
  \bibinfo{author}{\bibfnamefont{D.~F.} \bibnamefont{Shapiro}},
  \bibinfo{author}{\bibfnamefont{H.~G.} \bibnamefont{Craighead}},
  \bibnamefont{and} \bibinfo{author}{\bibfnamefont{J.~M.}
  \bibnamefont{Parpia}}, \bibinfo{journal}{Nano Letters}
  \textbf{\bibinfo{volume}{7}}, \bibinfo{pages}{1728} (\bibinfo{year}{2007}).

\bibitem[{\citenamefont{Cimalla et~al.}(2006)\citenamefont{Cimalla, Foerster,
  Will, Tonisch, Brueckner, Stephan, Hein, Ambacher, and
  Aperathitis}}]{cimallaAPL2006}
\bibinfo{author}{\bibfnamefont{V.}~\bibnamefont{Cimalla}},
  \bibinfo{author}{\bibfnamefont{C.}~\bibnamefont{Foerster}},
  \bibinfo{author}{\bibfnamefont{F.}~\bibnamefont{Will}},
  \bibinfo{author}{\bibfnamefont{K.}~\bibnamefont{Tonisch}},
  \bibinfo{author}{\bibfnamefont{K.}~\bibnamefont{Brueckner}},
  \bibinfo{author}{\bibfnamefont{R.}~\bibnamefont{Stephan}},
  \bibinfo{author}{\bibfnamefont{M.~E.} \bibnamefont{Hein}},
  \bibinfo{author}{\bibfnamefont{O.}~\bibnamefont{Ambacher}}, \bibnamefont{and}
  \bibinfo{author}{\bibfnamefont{E.}~\bibnamefont{Aperathitis}},
  \bibinfo{journal}{Applied Physics Letters} \textbf{\bibinfo{volume}{88}},
  \bibinfo{pages}{253501} (\bibinfo{year}{2006}).

\bibitem[{\citenamefont{Jiang et~al.}(2004)\citenamefont{Jiang, Yu, Liu, and
  Huang}}]{jiangPRL2004}
\bibinfo{author}{\bibfnamefont{H.}~\bibnamefont{Jiang}},
  \bibinfo{author}{\bibfnamefont{M.~F.} \bibnamefont{Yu}},
  \bibinfo{author}{\bibfnamefont{B.}~\bibnamefont{Liu}}, \bibnamefont{and}
  \bibinfo{author}{\bibfnamefont{Y.}~\bibnamefont{Huang}},
  \bibinfo{journal}{Physical Review Letters} \textbf{\bibinfo{volume}{93}},
  \bibinfo{pages}{185501} (\bibinfo{year}{2004}).

\bibitem[{\citenamefont{Mohanty et~al.}(2002)\citenamefont{Mohanty, Harrington,
  Ekinci, Yang, Murphy, and Roukes}}]{mohantyPRB2002}
\bibinfo{author}{\bibfnamefont{P.}~\bibnamefont{Mohanty}},
  \bibinfo{author}{\bibfnamefont{D.~A.} \bibnamefont{Harrington}},
  \bibinfo{author}{\bibfnamefont{K.~L.} \bibnamefont{Ekinci}},
  \bibinfo{author}{\bibfnamefont{Y.~T.} \bibnamefont{Yang}},
  \bibinfo{author}{\bibfnamefont{M.~J.} \bibnamefont{Murphy}},
  \bibnamefont{and} \bibinfo{author}{\bibfnamefont{M.~L.}
  \bibnamefont{Roukes}}, \bibinfo{journal}{Physical Review B}
  \textbf{\bibinfo{volume}{66}}, \bibinfo{pages}{085416}
  (\bibinfo{year}{2002}).

\end{thebibliography}
\end{document}